# Active-carbon based supercapacitors with Au colloids: the case for placing the colloids at the electrolyte/electrode interface

H. Grebel[(1)], Shupei Yu[(2)] and Yuanwei Zhang[(2)]

[(1)] Center for Energy Efficiency, Resilience and Innovation (CEERI) and the ECE Department at the New Jersey Institute of Technology, Newark, NJ 07102. grebel@njit.edu

[(2)] Department of Chemistry and Environmental Science at the New Jersey Institute of Technology, Newark, NJ 07102. yuanwei.zhang@njit.edu

**Abstract:**

Supercapacitors (S-C) are short-term energy storage elements that find many applications, e.g., electronic charging devices and suppressors of power fluctuations in grids that are interfaced with sustainable sources. The capacitance of an ordinary capacitor increases when dispersing metallic colloids in its dielectric. A similar strategy for S-C means a deployment of nano-scale metal colloids (in our case, Au nano particles, or AuNPs) at the very narrow interface between an electrolyte and the porous electrode (here, active-carbon film on a grafoil current collector). This is achieved by making the ligand that is coating the AuNPs negatively charged. We demonstrated a very large specific capacitance increase with a minute addition of functionalized AuNPs to the slurry. For example, C-V data at a scan rate of 20 mV/s indicated a specific capacitance amplification by a factor of 10 when 30 μg of AuNPs were incorporated with 200 mg of active carbon while using a 1 M $Na_2SO_4$ electrolyte and a 5% cellulose acetate butyrate as a binder. We make the case that the adhesion of the AuNPs to the surface of the electrode was strong: upon replacing the electrolyte, from 1 M $Na_2SO_4$ to 1 M KOH and retaining the same set of electrodes, the enhancement capacitance factor decreased as compared to 1 M $Na_2SO_4$ electrolyte but remained large, ~3, as determined by C-V traces at the same scan rate of 20 mV/s.

# I. Introduction

Originally devised for microwave lenses, Artificial Dielectrics (AD) are man-made materials that contain metal features much smaller than a characteristic propagating wavelength (e.g., sub-cm size ball bearings embedded in a dielectric and using a microwave wavelength of 3 cm to interrogate them) [1-3]. These features alter the effective permittivity and permeability of the dielectric through creation of locally induced dipoles. Examples are: a microwave lens [1] and nano-size semiconductor embedded dielectrics [4,5]. While there is a large body of work on AD at the high-frequency range (of the order of GHz, or even in the visible range), little if at all, has been investigated at the low frequency end (of the order of Hz, or kHz).

Supercapacitors (S-C) take advantage of the large capacitance at the narrow interface between a porous electrode and an electrolyte [6-11]. Here, we concentrate on carbon-based S-C that exhibit electrical double-layer. S-C are used in a wide-range of applications, such as consumer electronic products, memory back-up devices, hybrid electric vehicles, power supply system [12-13]. They were also proposed as buffers to highly fluctuating power grids that are equipped with sustainable sources [14]. Our intent is to gain basic knowledge on the capacitance changes when incorporating very low dispersion of gold colloids at the very narrow interface region between an electrolyte and a porous electrode; the latter is made of active-carbon (A-C) film deposited on a current-collector, made of grafoil. One ought to note that the Au colloids are not placed within the A-C particulate, but rather on its surface. Placing the AuNPs inside the A-C particulate would result only in an impedance change of the electrode itself with little effect on the cell capacitance.

AD may be understood in terms of local-fields theory [2]. In a quasi-DC approach, one analyzes the effect of a single colloid per unit cell, which is made of two parallel conductors. Without the colloids, the electrode's charge is Q under a cell bias of $V_{in}$. Under a cell bias, the metal colloids act as dipoles (namely, the free electrons in the metal are displaced) and the colloids become polarized. This in turn induce additional charge, q, on the cell's electrodes. For a given bias, the increased charge is translated into an increased capacitance. The unit cell is defined by the colloid spacing, a. The cell's capacitance is defined as $C_0$ without the colloid and C' with it. Assume that the colloid is spherical with a volume of $d^3$; the cell dimension is $a^2b$ with an electrode spacing of b; and that the external potential $V_{in}$ is expressed by the external field as, $E_0 \cdot b$ [2]. One may assess the effective new capacitance in the presence of colloids, as,

$$C' = C_0(1+q/Q) = C_0(1+C_{coll.}V_{coll.}/C_0V_{in}) \sim C_0\{1+[(d^2/d) \times (E_0 \cdot d)]/[(a^2/b) \times (E_0 \cdot b)]\} = C_0(1+N \cdot d^2). \qquad (1)$$

The last step is achieved when the lateral area of a unit cell is inversely proportional to the number of colloids, $N=1/a^2$ and there is one colloid per cell.

When adding more and more colloids, the local field at a particular colloid position is a linear combination of the external field and an interaction field. The local interaction field on a particular colloid is exerted by all other colloids excluding the colloid itself. The effect of all other colloids is characterized by an interaction parameter, C. At very low frequencies, the interaction field is mostly of electric nature. The polarization is summarized in a self-consistent form, the Clausius-Mossotti relation when the distance between colloids is much smaller than the wavelength of interest; this is easily achievable at low frequencies:

$$p = N\alpha_e E_0/(1-\alpha_e C). \qquad (2)$$

Here: N is the concentration of the metallic colloids, $\alpha_e$ is the electrical polarizability of the colloid and $E_0$ is the external field, applied by two parallel conductors. The relative, effective dielectric constant in comparison to the background material is derived from the Eq. 2 as,

$$k = 1 + N\alpha_e/(1-\alpha_e C). \qquad (3)$$

Thus, capacitance increase, which is related to the increase in the effective dielectric value is achieved with an increased number of dipoles and the interaction between them.

The surface of the porous electrode is a few orders of magnitude larger than its geometrical flat surface projection. It is of fractal nature [15-16] with a typical dimensionality of 2.5. Several methods have been devised to increase its effect [17-20]. The surface of the porous electrode is proportional to the geometrical volume of the sample, so it makes sense to normalize the capacitance results by the electrode volume for proper comparisons [21]. In our case, volumetric specific capacitance is amount to gravitonic specific capacitance because the area of all of our samples was the same.

Dispersions of the Au nano-particles (AuNPs) have been known for a long time. Their preparation methods are well-established, as well as, the relationship between their size, as determined by SEM or TEM and their optical scattering properties [22-23]. In order to ensure a good suspension, the colloids are functionalized with a negatively charged ligand. While the plasmonic peak absorption of uncoated AuNPs dramatically changes as a function of particles size [24], it is much less so for ligand coated particles [25]. A well-established technique that correlates the particle size to its optical scattering is dynamic light scattering (DLS) [26], which is used to determine the average particle size. The electrode itself is charge-neutral; nonetheless, there is an electrostatic bond between the neutral, yet conductive A-C electrode and the negatively functionalized AuNPs. We use this bond to attach the AuNPs to the A-C electrode.

The paper is organized as follows:
In Section II, Materials and Methods we describe the preparation and characterizations of the AuNPs, the titration procedure, the fabrication of the porous A-C electrodes, initial results and the case for volumetric normalization. In Section III, Results we describe experimental results on the various structures and related simulations. The Discussion is provided in Section IV and we make concluding remarks in Section V.

## II. Materials and Methods

### II.a Preparation methods - gold colloids:

Nano-size AuNPs were synthesized by following a well-established method [27]; it employs citrate as a reducing agent and stabilizer. In brief, chloroauric acid ($HAuCl_4$) water solution (10 mg $HAuCl_4$ in 90 mL of water) was heated to boiling and sodium citrate solution (0.5 mL of 250 mM) was introduced. The mixture was stirred for 30 min until the color turned to wine red, or purple-brown. AuNPs were then purified by centrifuge and washed with DI water three times. The concentration of AuNPs in water was 1 mg/mL; the titration experiments were made with increasing amounts of 10 $\mu$L per batch and are translated to $\mu$g when referenced to the amount of A-C in the batch; otherwise, they are quoted in $\mu$L.

There were 2 batches of suspended AuNPs in water: medium size of ca 45 nm and aggregates of ca 100 nm and whose normalized light absorption curves are shown in Fig. 1a. As can be seen from Fig. 1a, the peak position for the aggregates is red-shifted a bit compared to the smaller particles, is much broader and exhibits a significantly slower spectral decaying tail. The corresponding DLS curves are shown in Fig. 1b indicating the particles' diameter distribution. The Zeta potentials of the coated AuNPs were, -31.1 mV and -43.9 mV for the 45 nm and 100 nm colloids, respectively.

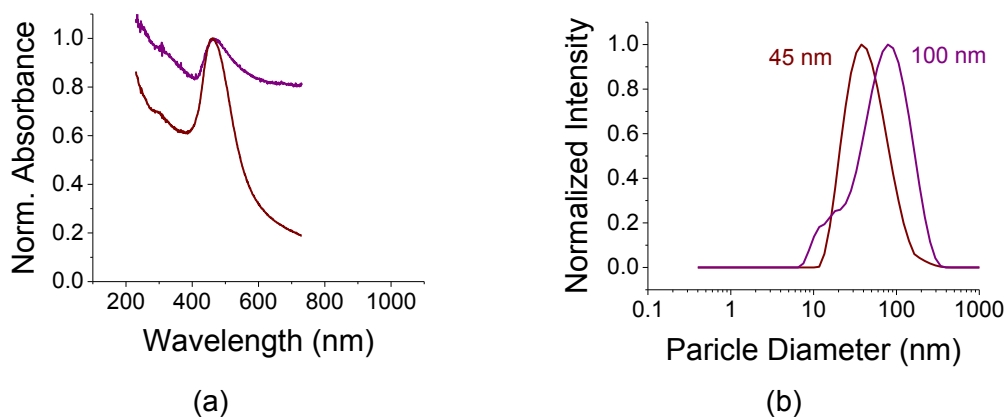

Figure. 1. (a) Absorbance (normalized such that the peak equals unity) for ligand-coated, small size (wine color) and aggregates (purple color) of AuNPs. Both were suspended in water. The plasmonic absorption is clearly seen; the aggregate peak is a bit red-shifted and broader compared to small particles' peak. The increase in absorption towards the UV spectrum is attributed to an increased scattering. (b) The corresponding DLS data show the two particle sizes used in the experiments.

### II.b. Preparation methods - the porous electrodes:

(a) A 100 mg of Cellulose Acetate Butyrate binder (CAB, Aldrich Chemicals) were first dissolved in 20 mL of acetone. 2 g of active-carbon (A-C, specific surface area of 1100 m$^2$/g, produced by General Carbon Company, GCC, Paterson, NJ, USA) were added and sonicated for 1 hour using a horn antenna. Six vials, each containing 1 mL, and sometimes 2 mL of the slurry were prepared. To these, succession amounts of 10 μL of AuNPs, suspended in water were added. Each mixture was further sonicated with the horn antenna for additional 30 min. The slurry was dropped casted on grafoil electrodes (area of contact 1.27x1.27 cm$^2$, manufactured by Miseal and purchased through Amazon), baked on a hot plate at <90 °C and then soaked with an electrolyte (1 M of $Na_2SO_4$, NaCl, or KOH). A fiberglass filter (Whatmen 1851-055) was used as a membrane. The Au colloid concentration was 1 mg/mL. The Au colloids sizes were of order of ~45 nm (Fig. 1b).

(b) Similar to (a) but with Poly-Vinyl Alcohol (PVA provided by EastChem Labs) as a binder and dissolved in water; 1 M of NaCl as an electrolyte. The Au colloids sizes were of order of ~100 nm, exhibited some sedimentation and deemed as aggregates (Fig. 1b).

On the average, the A-C particulates' dimension were ca 15 microns [28-29] and their specific surface area was rated as 1100 m$^2$/g. Through mixing, the AuNPs would end up either on the A-C particulate, in the binder, or in the electrolyte. We found out that incorporating the AuNPs in the electrolyte did not yield any capacitance increase. This is due to screening by the ions in the electrolyte. Administrating the AuNPs in the polymeric binder before adding the A-C was not efficient, either. The more efficient way was to first mix the A-C with the polymer and only then add the AuNPs. Adding the AuNPs to a thin polymeric layer directly on the grafoil film resulted in a very poor capacitor because it blocked the collection of current.

### II.c. The samples:

Cuts of 200 micron thick grafoil electrodes with back adhesive (1.27 cm x 2.54 cm) were placed on similar cut microscope slides. Before placing it on the slides, the grafoil electrodes were placed in 1 M NaOH container and were exposed to 30 sec of microwave radiation in a microwave oven to improve adhesion between the binder materials to the grafoil. The two slides were held by tweezers (or plastic clips) and the boundaries of the sample were left unsealed while soaking it in the electrolyte. The sample configuration is shown in Fig. 2a and its picture in Fig.2b.

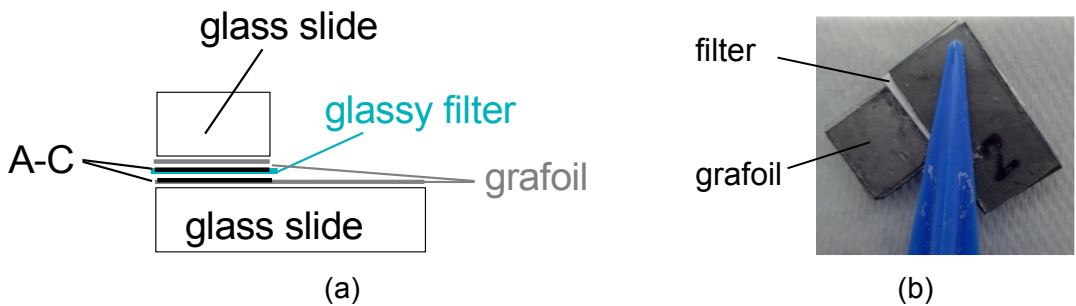

Figure 2. (a) A cross section of sample with grafoil coated active-carbon (A-C) electrodes, and (b) a picture of the cover area ca, 1.27x1.27 cm$^2$.

### II.d. Electrochemical Techniques:

Potentiostat/Galvanostat (Metrohm) was used in a 2-electrode set-up. Each sample was tested using three electrochemical methods: Cyclic Voltammetry (C-V) at scan rates of 20, 50 and 100 mV/s, Charge-Discharge (C-D) at applied currents of 0.5, 1 and 2 mA (translated to current densities of 0.31, 0.62 and 1.24 mA/cm$^2$) and Electrochemical Impedance Spectroscopy (EIS) between 50 kHz to 50 mHz. The results all agreed on the titration trends.

### II.e. Initial Characterizations – the effect of the current collectors and light:

A cell interfaced with un-coated 200 μm thick grafoil electrodes exhibited very small capacitance values compared to an A-C coated cell (Fig. 3a). The capacitive effect of the grafoil only electrode is less than 5% of the A-C coated sample. . Titration of AuNPs, drop-casted directly on grafoil-only electrodes exhibited little capacitance variations as a function of colloid concentration (Fig. 3b). The capacitance values were much smaller than those achieved when the AuNPs were interfaced with the A-C films (see below). One may, therefore conclude that the effects of the bare grafoil and AuNPs coated, grafoil (though without the A-C film) are minimal. The concentration of the AuNPs was referenced in this case to its suspension in 1 mL of acetone and the electrolyte was 1 M NaCl.

The AuNPs exhibited small response to optical radiation when exposed to a continuous white-light source. The A-C films were depositing transparent and conductive current collectors (ITO) similarly to [28-29]. Electrochemical Impedance Spectroscopy (EIS) shown in Fig. 3c revealed that there is a little change in the differential capacitance when the AuNP interfaced A-C films were exposed to the 75 W incandescent light. Overall, the effective electrode resistance (the real impedance value at the knee of the curve; see also Ref 30) has changed by only 6%. This change is attributed to optical excitations in the A-C film itself, the temperature change and to a lesser extent to the optical effect of the ITO films [28]. The corresponding C-D and C-V traces also exhibit a small change in the cell capacitance under the light exposure (Figs 3.d-e).

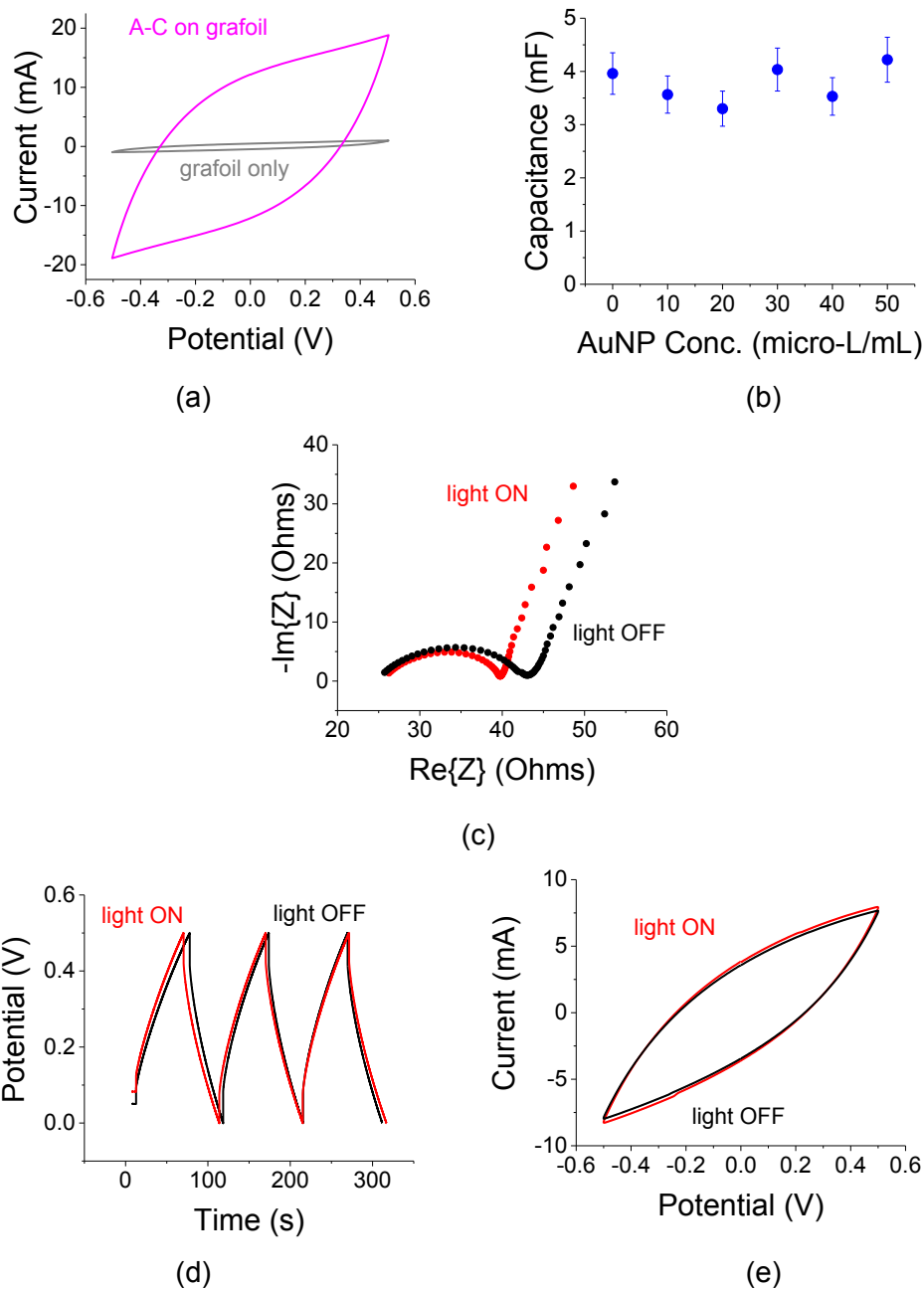

Figure 3. (a) C-V traces for A-C coated (pink curve), and uncoated (gray curve) a current collector made of bare grafoil. (b) Low concentration of AuNPs on grafoil only electrodes exhibited little capacitance variations. The concentration of the AuNPs is referenced to the solvent: 1 mL of acetone. The electrolyte was 1 M NaCl. (c) EIS curves under a continuous 75 W white-light exposure for colloidal embedded sample on ITO transparent substrates. (d) Corresponding C-D traces at $I_0$=1 mA (charge density, $J_0$=0.62 mA/cm$^2$) and (e) C-V traces at a scan rate of 100 mV/s revealed small light induced effects.

## II.f. Titration Experiments:

Titration experiments have been carried out to study the effect of AuNPs loading. The initial idea was that larger AuNP concentration will lend itself to larger capacitance values as more colloids will find itself at the electrolyte/electrode interface. Initial C-V experiments with 1 M of $Na_2SO_4$ were made as shown in Fig. 4a. These experiments exhibited a capacitance peak at AuNPs concentration of 40 μg/mL. The related capacitances as a function of AuNPs concentration are shown in Fig. 4b. Note a capacitance increase by a factor of ~2 as a result of AuNPs presence when compared to a reference sample without them (the sample with 0 AuNP concentration). The amount of active-carbon in each vial was 100 mg/mL.

The area of each sample was the same and the related capacitance values of Fig. 4a may be viewed as specific areal capacitance. For examples, the peak capacitance of 110 mF at 40 μg/100 mg is translated to 110 mF/$1.27^2$ $cm^2$=68.2 mF/$cm^2$. The trend in Fig. 4 is also exhibited in Fig.5a,b and Fig. 6a,b for C-D ($I_0$=1 mA, $J_0$=0.62 mA/cm2) and EIS experiments, respectively.

Two reference experiments were additionally made: (1) a repeat of the 50 μg experiment on another grafoil substrate to assess variations in sample preparation. That variation was estimated at ±5%. (2) 30 μg of AuNPs were added to the mixture holding 30 μL of AuNPs (in total 60 μL of AuNPs in 1 mL) to assess an increase in colloid concentration.

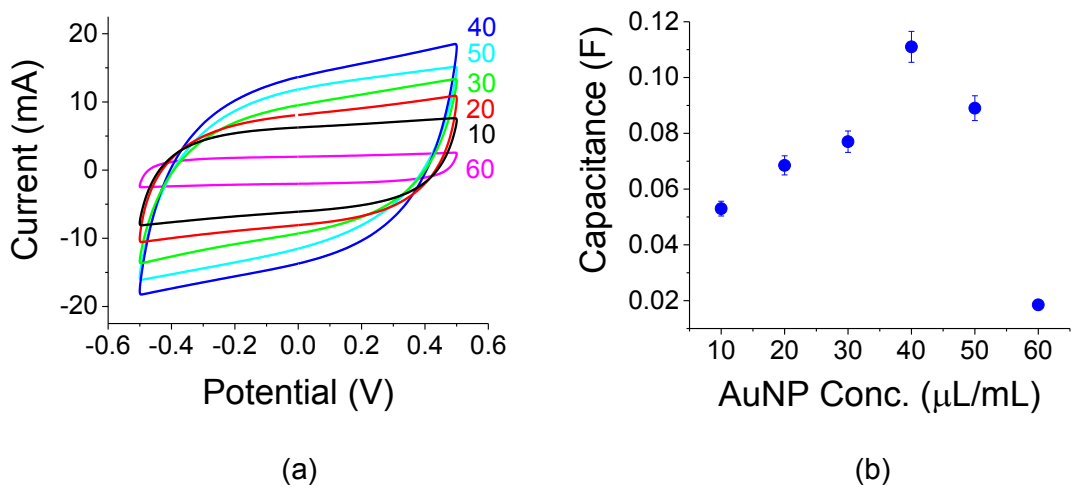

(a)              (b)

Figure 4. (a) Successive C-V plots for various concentrations (in μL per mL of solvent) depicted on the right. (b) Capacitance as a function of AuNPs concentration exhibits a peak at 40 μL/mL. The electrolyte was 1 M of $Na_2SO_4$.

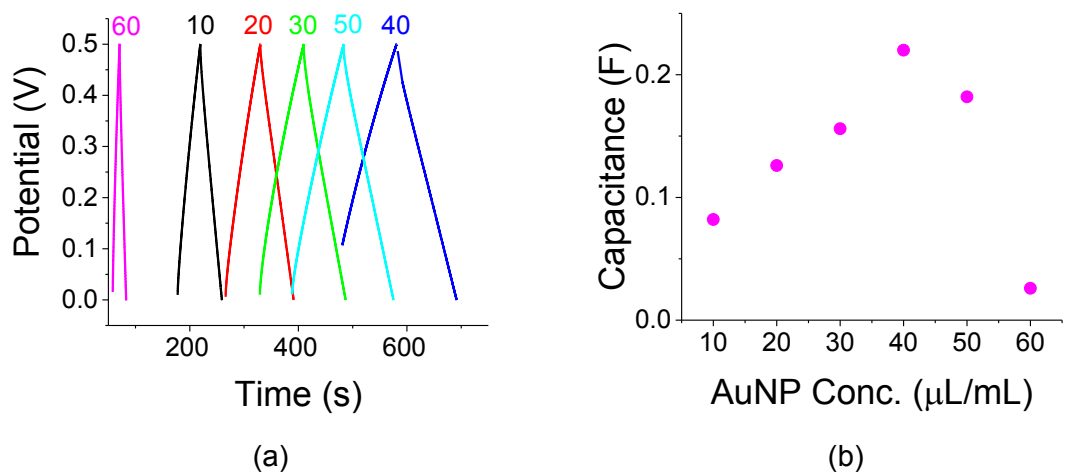

(a) (b)

Figure 5. (a) C-D plots for various AuNPs concentration (in μL per 1 mL of solvent). Current used was $I_0$=1 mA (charge density, $J_0$=0.62 mA/cm$^2$). (b) Capacitance as a function of AuNPs concentration exhibits a peak at 40 μL/mL. The electrolyte was 1 M of $Na_2SO_4$.

Electrochemical Impedance Spectroscopy (EIS) as a function of AuNPs concentration is shown in Fig. 6a. The capacitance value was derived from the slope of the low-frequency points in the -Im{Z} vs 1/frequency curve as shown in Fig. 6b for the peak capacitance at 40 μL/mL. The slope of the curve is proportional to the inverse of the differential capacitance, $C_{diff}$, according to $1/(2\pi C_{diff})$ [30]. From Fig. 6b, $C_{diff}$ is assessed as 194 mF, corroborating the value obtained from Fig. 2b. The electrode resistance (ca 3-5 Ohms for all samples) was rather small.

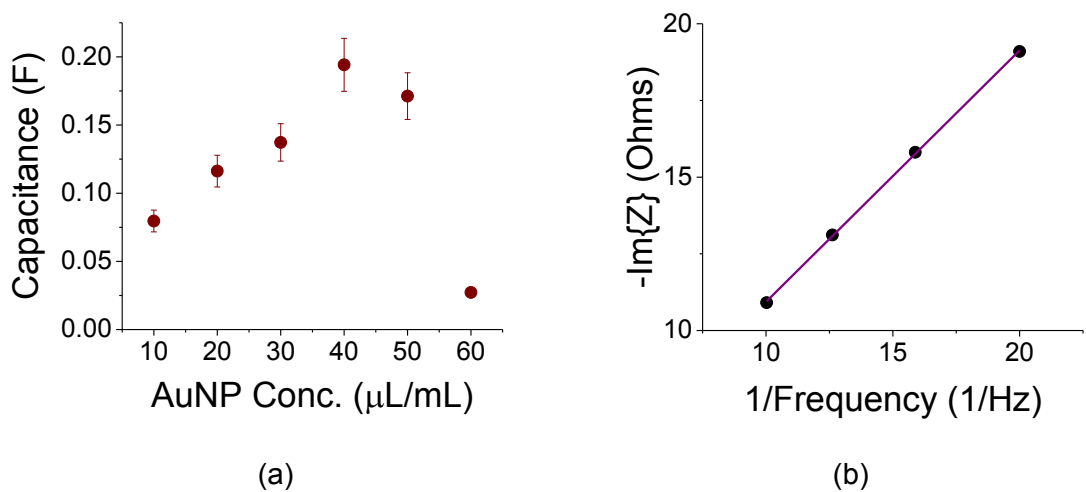

(a) (b)

Figure 6. (a) Capacitance values, derived from EIS curves. The concentration of the AuNP is in mL per 1 μL of solvent. (b) The slope of the linear curve for the sample exhibiting capacitance maxima at 40 μL/mL of AuNPs is proportional to $1/C_{diff}$. The electrolyte was 1 M of $Na_2SO_4$.

One could argue that values of film capacitance, or even values of specific capacitance with respect to the area (in units of [F/cm$^2$]), do not provide for a full picture; as pointed earlier, the effective (porous) film's surface scales as the sample's volume. Plots of the specific gravitonic capacitance in units of F/g were made for different set of films at various conditions (various thicknesses, binders and electrolytes) as outlined below. As also outlined below, the gravitonic capacitance is directly related to the volumetric capacitance in our case:

The area of each sample was the same, 1.61 cm$^2$. The density of the composite film, $d_{film}$, was assessed as the weighted density of its components: 95% of A-C and 5% of polymeric binder. It was kept the same throughout all tests: $d_{film}$=0.95×($d_{A-C}$=0.375 g/cm$^3$) + 0.05×($d_{polymer}$=1.21 g/cm$^3$) ~ 0.42 g/cm$^3$. The volume was assessed as A×t and the film thickness was t=$w_{film}$/($d_{film}$×A) with $w_{film}$ being the film weight. The gravitonic specific capacitance in units of F/g is related to the volumetric specific capacitance in units of [F/cm$^3$] as: [F/cm$^3$]=[F×(g/cm$^3$)/g]=[F×$d_{film}$/g]. Therefore, gravitonic and volumetric specific capacitance values are proportional to one another through a constant: the film density, or, $d_{film}$=0.42 g/cm$^3$.

## III. Results

### III.a. Thin and Thick Samples:

Two sets of experiments were first carried out: one for thicker (~450 μm, electrode loading of 14 mg/cm$^2$) and the other for thinner (~75 μm, electrode loading of 6 mg/cm$^2$) films. Their results are shown in Fig. 7. The thin films were obtained by further dilution of the original batches with 1 + 1 mL of acetone + ethanol for a total of 3 mL solution. The concentration of the AuNPs is referenced to the amount of the A-C in the sample because both the AuNPs and the A-C were diluted by the same amount. The trend exhibited by the thicker and thinner films is the same, though thinner films exhibited larger specific values [31]. The consistent results for the reference sample (0 μg/100 mg) alludes to the repeatability of the experiments (note that the points for 10 μg/100 mg coincide with each other). Of interest are the relatively large values at 20 and 50 μg/100mg. The large variability of values at 50 μg/100 mg in Fig. 7 may be attributed to lower film integrity due to a mixture of hydrophobic and hydrophilic components; the AuNPs are suspended in water while the A-C and the CAB binder are hydrophobic.

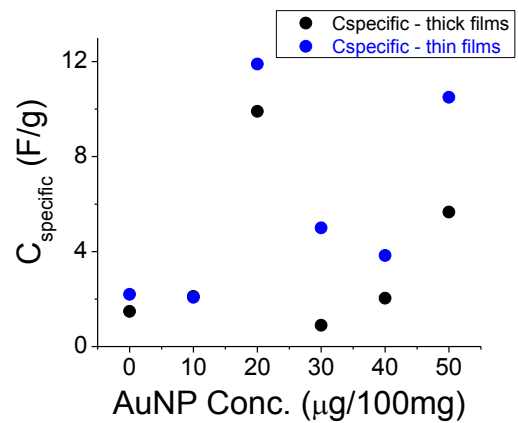

Figure 7. Specific capacitance, $C_{specific}$ in units of F/g from C-V data for thick and thin films with a CAB binder. The concentration of AuNPs is referenced to the amount of A-C in the slurry. The $C_{specific}$ value for 10 μg per 100 mg of A-C is identical for both films. The electrolyte was 1 M of Na$_2$SO$_4$.

Another set of tests was conducted with the smaller size AuNPs colloids (~45 nm of Fig. 1a), yet with 1 M NaCl as an electrolyte. Better film homogeneity was achieved by first sonicating the AuNPs with A-C in 1 mL of acetone following with 1 mL of 6% CAB by weight in acetone, as well; the latter mixture was re-sonicated. The results shown in Fig. 8 are consistent with the previous results. The peak value seem to shift a bit towards a larger concentration.

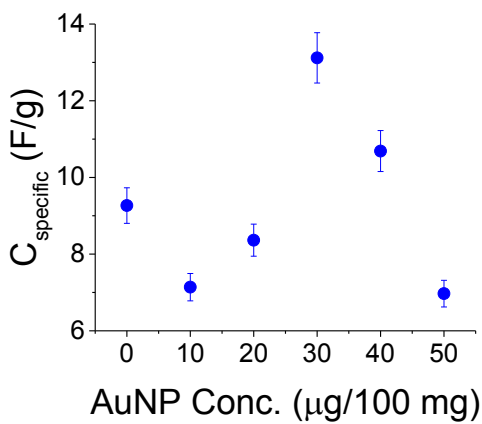

Figure 8. Specific capacitance values derived from C-V data for CAB binder and 1 M NaCl electrolyte. Similar trend is obtained with C-D data. The concentration of AuNP in $\mu$g is referenced to 100 mg of A-C in the batch.

Another set of experiments was conducted with a PVA binder and a larger amount of A-C (200 mg per 1 mL of water). These experiments were conducted with the larger size AuNP (~100 nm). Unlike the hydrophobic CAB, PVA is hydrophilic and better matches with the water suspended AuNPs as alluded to by [32]. PVA is less compatible with the hydrophobic current collector and the treatment with the NaOH helped the adhesion of the film to the grafoil. The electrolyte here was 1 M NaCl. Two sets of experiments were conducted and their averages are shown in Fig. 9 by the blue dots. The data are consistent up to 40 $\mu$g per 200 mg of A-C after which film homogeneity due to charged AuNPs packing could become an issue.

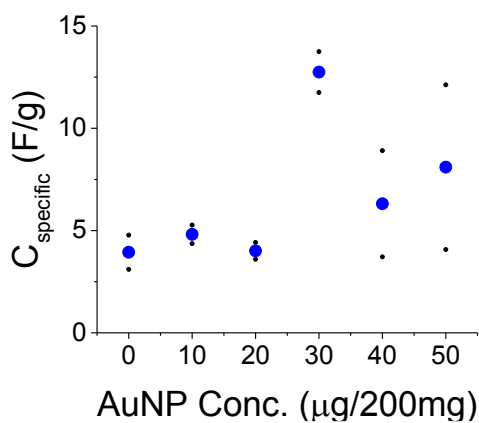

Figure 9. Specific capacitance values derived from C-V data for PVA binder and 1 M NaCl electrolyte. The black dots are the results from two sets of experiments; the blue dots are averages of these two experiments. The concentration of AuNP in $\mu$g is referenced to 200 mg of A-C in the batch. Large fluctuations above 40 $\mu$g per 200 mg of A-C are result of charged AuNPs packing.

Complementary C-D and C-V data with a CAB binder and 1 M $Na_2SO_4$ electrolyte are shown in Figs. 10 and 11, respectively for a reference and a 30 μg of AuNPs per 200 mg of A-C. The C-D data revealed little change as a function of the applied current, $I_0$. The enhancement factor due to the presence of the AuNPs is approximately 6.

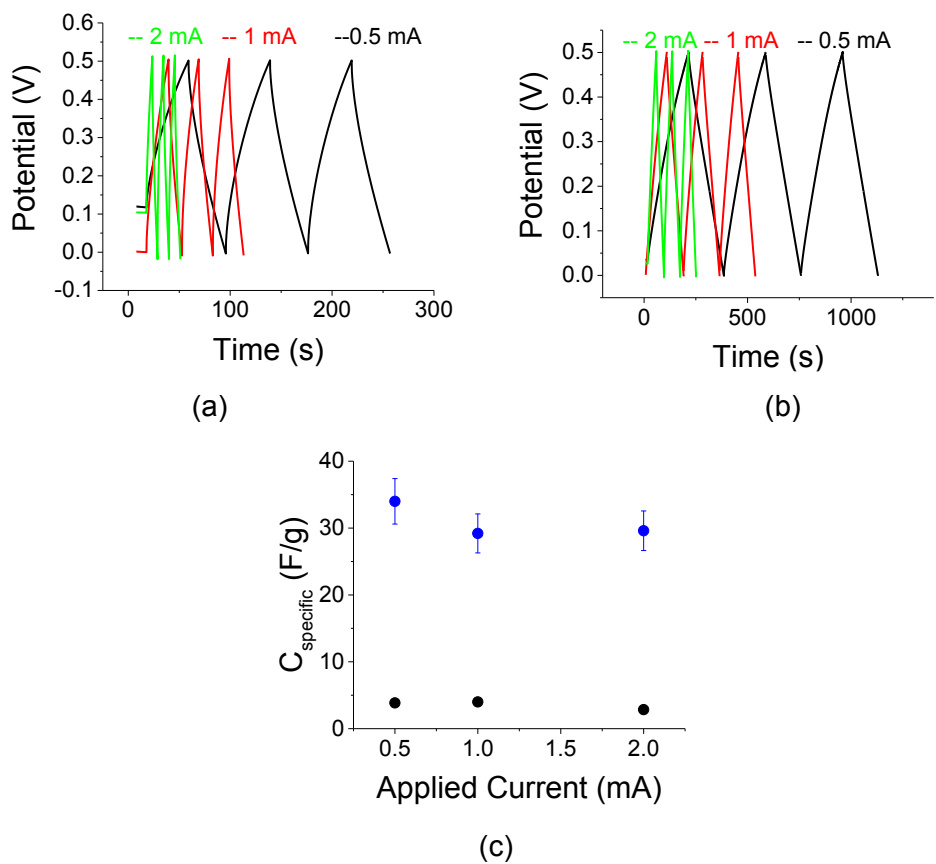

Figure 10: C-D data as a function of various current levels: $I_0$=0.5, 1 and 2 mA ($J_0$=0.31, 0.62 and 1.24 mA/cm², respectively) for: (a) reference sample (no AuNPs) and (b) with 30 μg of AuNPs per 200 mg of A-C. (c) The gravitonic specific capacitance as a function of the applied current for the AuNPs sample (blue dots) and the reference (black dots).

C-V curves are shown in Fig. 11 for the sample of Fig. 10. As indicated by the inclination of the curve, the resistance of the sample decreases upon decreasing scan rates (Figs. 11 a-b). This information is also conveyed by the specific capacitance plot of Fig. 11c. The specific capacitance typically depends on the C-V scan rate due to the effect of electrolyte diffusion. The enhancement factor at a scan rate of 20 mV/s is more than 10-fold.

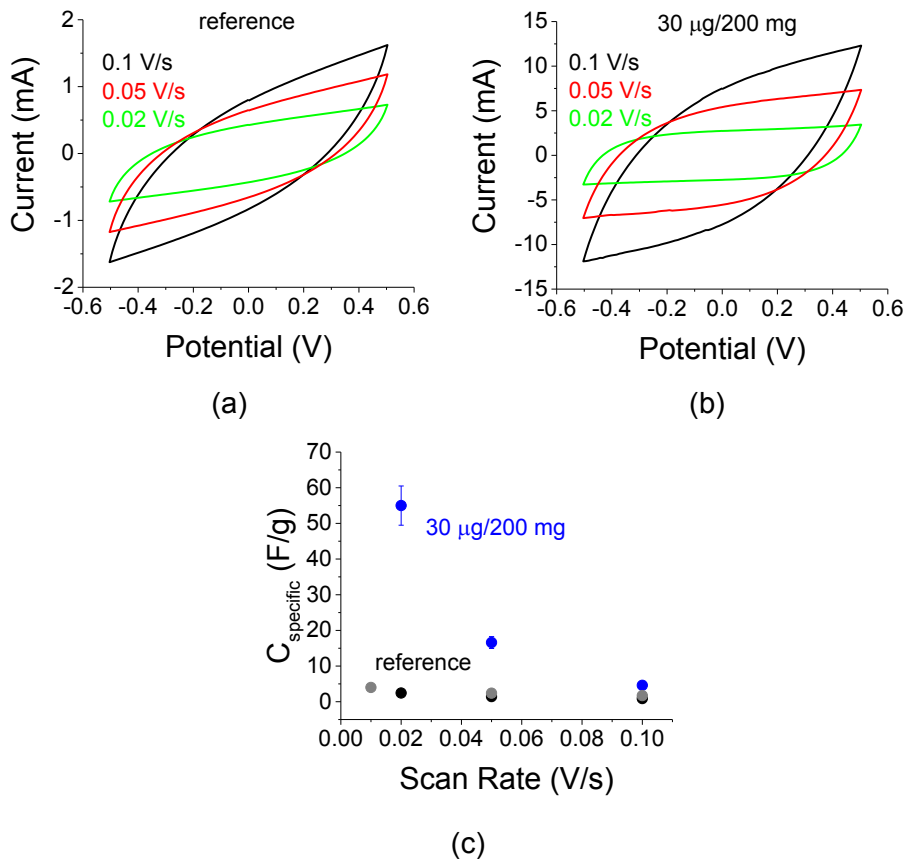

Figure 11: C-V data as a function of various scan rates for (a) reference sample (no AuNPs) and (b) with 30 µg of AuNPs per 200 mg of A-C in a 1 M $Na_2SO_4$ electrolyte. (c) The gravitonic specific capacitance as a function of the scan rate. The experiment for the reference sample was repeated twice (the gray and black dots).

Multiple C-V scans were conducted on the 30 µg of AuNPs per 200 mg of A-C sample (Fig. 12). The open-edge sample (Fig. 2b) was immersed in a $Na_2SO_4$ electrolyte while the contacts to the current collectors were above the electrolyte level. This way, one avoids drying due to inappropriate sealing. The change between the first and last scan was less than 2%, exhibiting a small capacitance increase.

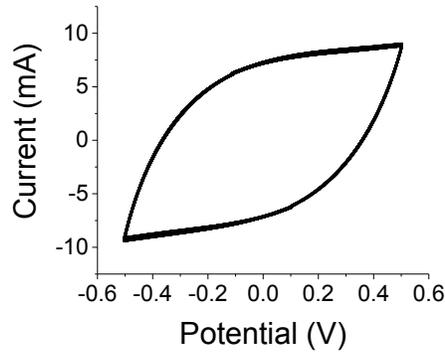

Figure 12: 125 C-V scans for the 30 μg of AuNPs per 200 mg of A-C sample in 1 M $Na_2SO_4$ at a rate of 50 mV/s exhibited <2% change between the first and last scan; .

Finally, we replaced the 1 M $Na_2SO_4$ electrolyte with 1 M KOH (see also Refs 33-35). The open edge sample was thoroughly washed with water and the glassy filter was replaced with no visible damage to the electrodes. The C-V and C-D traces in Fig. 13 indicate a large enhancement factor of ca 3 and 4.5, respectively for the specific capacitance. Indirectly, the experiments also point to the integrity of the electrodes and the strength of the bond between the A-C and the AuNPs.

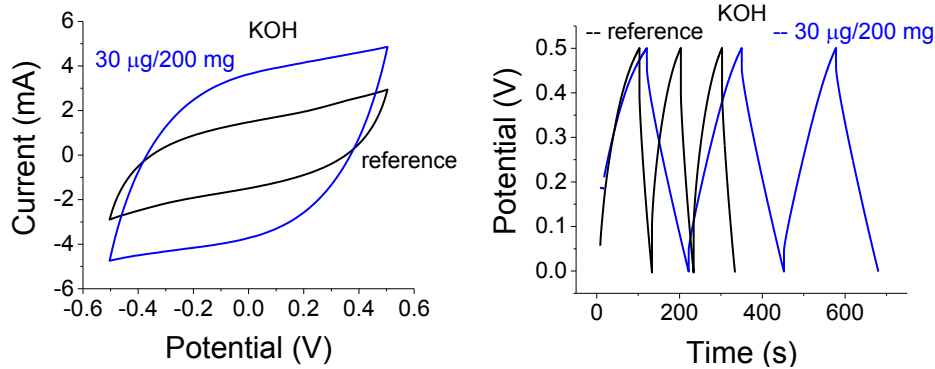

Figure 13: Replacing the 1 M $Na_2SO_4$ electrolyte by 1 M KOH while retaining the same electrodes. (a) C-V traces at a scan rate of 20 mV/s help assess the enhancement factor of the specific capacitance as ~3. (b) C-D traces at an applied current of $I_0$=1 mA (charge density, $J_0$=0.62 mA/cm$^2$) help assess the enhancement factor of the specific capacitance as ~4.5.

**Simulations:**

In the simulations, one compares two cases: (1) a cell without metal colloids; (2) a cell with colloids extruding into the neutral region. A portion of the S-C was modelled by a parallel plate capacitor with a graphitic film of finite conductance that is deposited on top of a metallic current collector. The gold particle is partially embedded inside the graphic film. The neutral region is situated at the cell's center and a bias is provided between the two current collectors: $\pm V_{in}$. The surface potential of the neutral region and the surface potential of the metal colloids are floating, and hence depend on the local field. As a result and as expected, the simulation indicates that neutral region maintains zero potential when the electrodes are biased (Fig. 14). The colloid radius was 10 nm (particle size or particle diameter of 20 nm). Larger particles would result in larger polarization effect. The otherwise capacitance increase is diminished when the colloids are embedded away from the electrode. This corroborates the experiments; when the AuNPs were mostly part of the electrolyte, the efficiency of the capacitor was substantially diminished. The conductivity of the graphitic material matters only if its conductance is comparable to the current collectors' value; this is generally not the case since the current collector is treated as a perfect conductor and the graphite-like material has a much smaller conductivity.

The local polarization in response to the external bias and is shown in Fig. 15. The latter points to a large polarization increase in the cell upon the presence of the AuNPs.

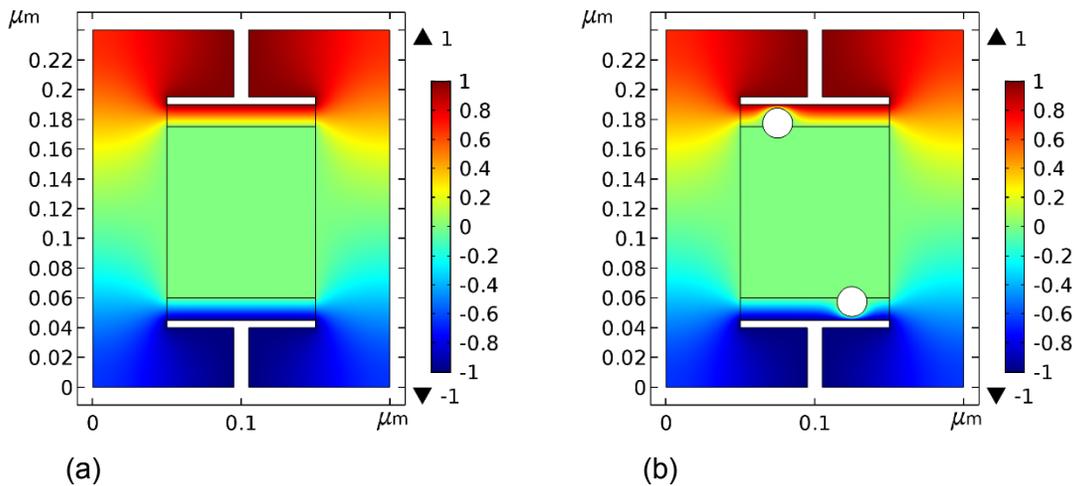

(a)      (b)

Figure 14. Potential distribution for $V_{in}$=1 V. The neutral region indeed has a constant potential (zero in this case) and the presence of the metallic colloid alters the potential lines. (a) No colloid. (b) With 10 nm colloids. The thermal legend is in units of Volts.

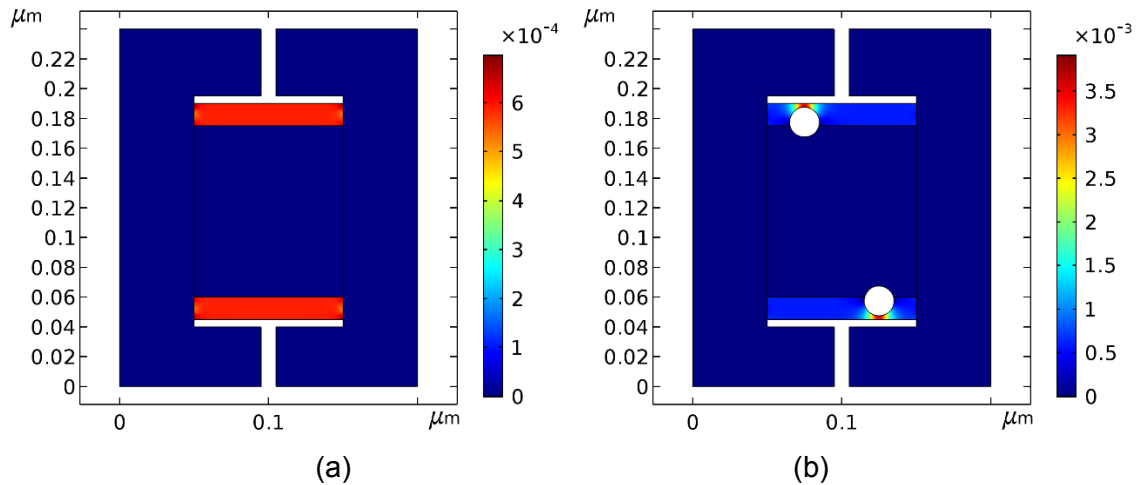

(a)      (b)

Figure 15. The absolute value of polarization is stronger near the colloids. The thermal legend is in units of C/m$^3$.

The capacitance of the cell is proportional to the charge on the electrode. Fig. 16 shows the change in the electrode charge as a function of the cell bias. The electrode charge, Q, is related to the bias, $V_{in}$, as, $Q = C \cdot V_{in}$, with C, the cell capacitance. The cell capacitance is therefore,

$C=Q/V_{in}$ and it is constant since the charge on the electrode varies linearly with $V_{in}$. The slope of the curve (the capacitance) increases upon presence of the AuNPs. The curves of Fig. 12 are translated to cell capacitance values of: 0.133 and 0.22 nF, for respectively, cells without colloids and cells with them.

While not shown, packing more colloids, say two AuNPs on each surface, obviously increases the cell capacitance but up to a point. The cell capacitance for very closely packed colloids on each surface is actually smaller than when they are placed apart. Simulation wise, the change is not large (ca 2.5%), yet noticeable and is attributed to a strong dipole-dipole interaction.

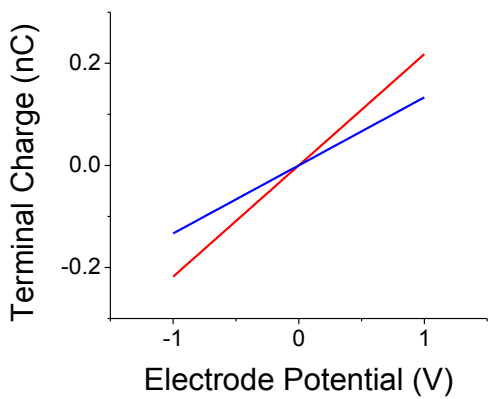

Figure 16. Change in electrode charge as a function of the bias, $V_{in}$. Blue line: graphite-like electrode without any colloid; Red line: graphite-like electrode with colloids on its surface. The cell capacitance is the slope of the linear Q-V curve.

In the introduction we made the case that the metallic colloids on the surface of an electrode may be viewed as electrical dipoles – the free electronic charge enables the polarization of the metal colloid under the external field between the surface of the electrode (the A-C film on the current collector) and the equally and opposite charge of the ions. The capacitance increase in this case was attributed to the formation of local dipoles and the interaction between them. Another point of view is to treat the colloids as small metallic capacitors. These are also polarized under the external bias. However, along with this view (a capacitor-within-capacitor, see Ref 36) the capacitance increase is attributed to the large polarization between the colloid and the nearest electrode surface as seen in Fig. 15b. Put it simply, the smaller the distance between the colloid and electrode, the larger is the effect.

## IV. Discussion

Experimentation and simulations exhibited enhancement of specific capacitance upon treating active-carbon electrodes with AuNPs. This effect occurred under various conditions (hydrophobic and hydrophilic polymeric binders, electrolytes and various AuNPs/A-C mass ratios). The effect was quite strong exhibiting enhancement factors of 10 despite the small mass of AuNPs with respect to the mass of A-C – a ratio of four orders of magnitude. In addition, the concentration used for the AuNPs is three orders of magnitude smaller than the concentration used for the optically excited yet neutral semiconductor colloids [28-29]. Optical excitations of the semiconductor particles made their oscillator strength similar to that of metallic colloids; yet, the semiconductor particles were dispersed throughout the entire film, including the binder. This observation in addition to the simulations point to the effective role of placing the metallic colloids within the double layer region, close to the A-C electrode.

It was also found out that the cell's capacitance was not affected by adding the colloids to the electrolyte, most likely due to ion screening. Bare current collector exhibited small capacitance variations with respect to the AuNPs concentration, which points to the important role played by the porous A-C films. An increase of A-C mass in the batch (100 mg/mL in Fig. 8 vs 200 mg/ml in Fig. 9) did not affect the presence of a large specific capacitance peak in our experiments. This suggests that the AuNPs were mostly adsorbed by the porous electrode regardless of its initial concentration levels. It also reaffirms the use of gravitonic, or volumetric specific capacitance because such normalization method factors out contributions from the film thickness. One takes advantage of the negatively charged ligand coating of the AuNPs, which enabled it to better adhere to the active-carbon electrode. Proof that the interaction between AuNPs and the A-C electrode was electrostatically strong was provided by Fig. 2 and 13. For Fig. 2, further dilution of the batch did not alter the titration trend; for Fig 13, replacing the electrolyte did not diminished the presence of an enhancement factor.

The ligand, coating the AuNPs enables a better colloid suspension. It may also limit the packing of the colloids on the sample's surface. One may argue that such effect would eventually lead to saturation in the value of the specific capacitance. At the same time, beyond the specific capacitance peak one may observe large specific capacitance fluctuations (e.g., Figs. 7, 9), which suggest that larger packing of AuNPs may increase the inhomogeneous colloid dispersion throughout the surface of the A-C electrode.

## V. Conclusion

Incorporating negatively charged, functionalized gold particles (at the level of tens of $\mu$g) with active-carbon (at the level of hundreds of mg) in aqueous based, supercapacitors exhibited a large specific capacitance enhancement (gravitonic, or volumetric) that can reach ten-fold when compared to its reference. Simulations point to the enhancing effect of AuNPs dipoles when placed at the electrode/electrolyte interface.

**References:**

1. Brown, John, and Willis Jackson. "The properties of artificial dielectrics at centimetre wavelengths." Proceedings of the IEE-Part B: Radio and Electronic Engineering 102.1 (1955): 11-16.

2. R. E. Collin, *Field Theory of Guided Waves*, Wiley-IEEE Press, 2nd edition, 1990.

3. Shih-Chang Wu and H Grebel, "Phase shifts in coplanar waveguides with patterned conductive top covers, J. Phys. D: Appl. Phys. 28 437-439 (1995)

4. H. Grebel and P. Chen, "Artificial dielectric polymeric waveguides: metallic embedded films ", JOSA A, 8, 615-618 (1991). https://doi.org/10.1364/JOSAA.8.000615

5. H. Grebel and P. Chen, "Artificial dielectric polymeric waveguides: semiconductor-embedded films", Opt. Letts, 15, 667-669 (1990). https://doi.org/10.1364/OL.15.000667

6. Keh-Chyun Tsay, Lei Zhang, Jiujun Zhang, Electrochimica Acta 60 (2012) 428–436.

7. Yudong Li, Xianzhu Xu, Yanzhen He, Yanqiu Jiang and Kaifeng Lin, Polymers 2017, 9, 2; doi:10.3390/polym9010002.

8. M. Kaempgen, C. K. Chan, J. Ma, Y. Cui, and G. Gruner, Nano Letts., , 9 (2009) 1872.

9. Michio Inagaki, Hidetaka Konno, Osamu Tanaike, Journal of Power Sources 195 (2010) 7880–7903.

10. Zhang, S and Pan, N, 2015. DOI:10.1002/aenm.201401401. https://escholarship.org/uc/item/26r5w8nc

11. Mohammad S. Rahmanifar, Maryam Hemmati, Abolhassan Noori, Maher F. El-Kady, Mir F. Mousavi, Richard B. Kaner, Materials Today Energy 12 (2019) 26-36. https://doi.org/10.1016/j.mtener.2018.12.006

12. B.K. Kim, S. Sy, A. Yu, J. Zhang, Electrochemical supercapacitors for energy storage and conversion, Handbook of Clean Energy Systems (2015) 1–25.

13. Xin Miao, Roberto Rojas-Cessa, Ahmed Mohamed and Haim Grebel, "The Digital Power Networks: Energy Dissemina-tion Through a Micro-Grid", Proceedings - IEEE 2018 International Congress on Cybermatics, 230-235. DOI: 10.1109/Cybermatics_2018.2018.00068

14. Roberto Rojas-Cessa, Haim Grebel, Zhengqi Jiang, Camila Fukuda, Henrique Pita, Tazima S. Chowdhury, Ziqian Dong and Yu Wan, "Integration of alternative energy sources into digital micro-grids", Environmental Progress & Sustainable Energy, (2018) 37, 155-164. DOI 10.1002/ep.

15. M. Jaroniec, R.K. Gilpin, J. Choma, "Correlation between microporosity and fractal dimension of active carbons", Carbon, 31, 1993, 325-331

16. F. E. Dolle, A. Lavancy, F. Stoeckll, Determination of the surface dimension of fractal dimension of active carbon by mercury porosimetry", J. Colloid and interface Science, (1994), 166, 451-461.

17. Li Feng, Bing Yan, Jiaojiao Zheng, Jiayun Chen, Rongyun Wei, Shaohua Jiang, Weisen Yang, Qian Zhang, Shuijian He, "Soybean protein-derived N, O co-doped porous carbon sheets for supercapacitor applications", New J. Chem., (2022)46, 10844-10853. https://doi.org/10.1039/D2NJ01355J

18. Si Zheng, Jianwei Zhang, Hongbing Deng, Yumin Du Xiaowen Shi, "Chitin derived nitrogen-doped porous carbons with ultrahigh specific surface area and tailored hierarchical porosity for high performance supercapacitors", Journal of Bio resources and Bio products, 6, (2021), 142-151. https://doi.org/10.1016/j.jobab.2021.02.002

19. Mohammad Reza Saeb, Navid Rabiee, Farzad Seidi, Bahareh Farasati Far, Mojtab Bagherzadeh, Eder C. Lima, Mohammad Rabiee, "Green $CoNi_2S_4$/porphyrin decorated carbon-based nanocomposites for genetic materials detection Journal of Bio resources and Bio products", 6, (2021), 215-222. https://doi.org/10.1016/j.jobab.2021.06.001

20. Cheng Du, Ping Li, Zhihua Zhuang, Zhongying Fang, Shuijian He, Ligang Feng, Wei Chen "Highly porous nanostructures: Rational fabrication and promising application in energy electrocatalysis", Coordination Chemistry Reviews, 466, (2022), 214604. https://doi.org/10.1016/j.ccr.2022.214604

21. Y. Gogotsi and P. Simon, "True Performance Metrics in Electrochemical Energy Storage", Science 334, 917 (2011). DOI: 10.1126/science.1213003

22. John Turkevich, "Colloidal Gold. Part I: HISTORICAL AND PREPARATIVE ASPECTS, MORPHOLOGY AND STRUCTURE", Gold Bull., 1985, 18, 86-91.

23. John Turkevich, " Colloidal Gold. Part II: COLOUR, COAGULATION, ADHESION, ALLOYING AND CATALYTIC PROPERTIES", Gold Bull., 1985, 18, 125-131.

24. Wolfgang Haiss, Nguyen T. K. Thanh, Jenny Aveyard, and David G. Fernig, "Determination of Size and Concentration of Gold Nanoparticles from UV-Vis Spectra", Anal. Chem. 2007, 79, 4215-4221.

25. Xiao-Dong Zhang, Di Wu, Xiu Shen, Pei-Xun Liu, Na Yang, Bin Zhao, Hao Zhang, Yuan-Ming Sun, Liang-An Zhang, Fei-Yue Fan "Size-dependent in vivo toxicity of PEG-coated gold nanoparticles", International Journal of Nanomedicine, 2011:6 2071–2081. http://dx.doi.org/10.2147/IJN.S21657

26. Jörg Stetefeld, Sean A. McKenna, Trushar R. Patel, "Dynamic light scattering: a practical guide and applications in biomedical sciences", Biophys Rev (2016) 8:409–427. DOI 10.1007/s12551-016-0218-6

27. Mercado-Lubo, R.; Zhang, Y.; Zhao, L.; Rossi, K.; Wu, X.; Zou, Y.; Castillo, A.; Leonard, J.; Bortell, R.; Greiner, D. L.; Shultz, L. D.; Han, G.; McCormick, B. A. "A Salmonella nanoparticle mimic overcomes multidrug resistance in tumours", Nature Communications 2016, 12225.

28  H. Grebel, " Asymmetric Supercapacitors: Optical and Thermal Effects When Active Carbon Electrodes Are Embedded with Nano-Scale Semiconductor Dots", C 2021, 7(1), 7; https://doi.org/10.3390/c70100072 .

29. H. Grebel, "Optically Controlled Supercapacitors: Functional Active Carbon Electrodes with Semiconductor Particles", Materials 2021, 14(15), 4183; https://doi.org/10.3390/ma14154183

30. Bing-Ang Mei, Obaidallah Munteshari, Jonathan Lau, Bruce Dunn, and Laurent Pilon, "Physical Interpretations of Nyquist Plots for EDLC Electrodes and Devices", J. Phys. Chem. C 2018, 122, 194−206. DOI: 10.1021/acs.jpcc.7b10582.

31. Suvi Lehtimäki, Anna Railanmaa, Jari Keskinen, Manu Kujala, Sampo Tuukkanen Donald Lupo, "Performance, stability and operation voltage optimization of screen-printed aqueous supercapacitors", Scientific Reports, 7:46001 (2017). DOI: 10.1038/srep46001

32. Qinxing Xie, Xiaolin Huang, Yufeng Zhang, Shihua Wu, Peng Zhao, "High performance aqueous symmetric supercapacitors based on advanced carbon electrodes and hydrophilic poly(vinylidene fluoride) porous separator", Applied Surface Science 443 (2018) 412–420.

33. Qian Zhang, Bing Yan, Li Feng, Jiaojiao Zheng, Bo You, Jiayun Chen, Xin Zhao, Chunmei Zhang, Shaohua Jiang and Shuijian He, "Progress in the use of organic potassium salts for the synthesis of porous carbon nanomaterials: microstructure engineering for advanced supercapacitors", Nanoscale, (2022) 14, 8216–8244. DOI: 10.1039/D2NR01986H.

34. Changshui Wang, Bing Yan, Jiaojiao Zheng, Li Feng, Zhenzhao Chen, Qian Zhang, Ting Liao, Jiayun Chen, Shaohua Jiang, Cheng Du, Shuijian He, "Recent progress in template-assisted synthesis of porous carbons for supercapacitors", Advanced Powder Materials, (2022) 1, 100018. https://doi.org/10.1016/j.apmate.2021.11.005.

35. Bing Yan, Jiaojiao Zheng, Feng Wang, Luying Zhao, Qian Zhang, Wenhui Xu, Shuijian He, "Review on porous carbon materials engineered by ZnO templates: Design, synthesis and capacitance performance", Materials and Design, (2021) 201, 109518. https://doi.org/10.1016/j.matdes.2021.109518

36. H. Grebel, "Capacitor-within-capacitor", SN Appl. Sci. 1, 48 (2019). https://doi.org/10.1007/s42452-018-0058-z